\documentclass[a4paper,fleqn,usenatbib]{mnras}
\usepackage[T1]{fontenc}
\usepackage{ae,aecompl}
\usepackage{graphicx}
\usepackage{amsmath,amssymb}
\usepackage{textgreek}
\newcommand{\abs}[1]{\lvert #1\rvert}
\newcommand{\ur}[1]{\,\mathrm{#1}}
\usepackage{siunitx}
\title[Line-Depth Ratio in $0.97$--$1.32\,\text{\textmugreek m}$]{Method to Estimate the Effective Temperatures of Late-Type Giants using Line-Depth Ratios in the Wavelength Range $0.97$--$1.32\,\text{\textmugreek m}$}
\author[D. Taniguchi et al.]{Daisuke Taniguchi,$^{1}$\thanks{E-mail: taniguchi@astron.s.u-tokyo.ac.jp} Noriyuki Matsunaga,$^{1,2}$\thanks{E-mail: matsunaga@astron.s.u-tokyo.ac.jp} Naoto Kobayashi,$^{3,4,2}$ Kei Fukue,$^{2}$
\newauthor Satoshi Hamano,$^{2}$ Yuji Ikeda,$^{2,5}$ Hideyo Kawakita,$^{2,6}$ Sohei Kondo,$^{2}$
\newauthor Hiroaki Sameshima$^{2}$ and Chikako Yasui$^{7}$ \\
$^{1}$Department of Astronomy, The University of Tokyo, 7-3-1 Hongo, Bunkyo-ku, Tokyo 113-0033, Japan \\
$^{2}$Laboratory of Infrared High-resolution spectroscopy (LiH), Koyama Astronomical Observatory, Kyoto Sangyo University, Motoyama, \\ Kamigamo, Kita-ku, Kyoto 603-8555, Japan \\
$^{3}$Institute of Astronomy, The University of Tokyo, 2-21-1 Osawa, Mitaka, Tokyo 181-0015, Japan \\
$^{4}$Kiso Observatory, The University of Tokyo, 10762-30 Mitake, Kiso-machi, Kiso-gun,Nagano 397-0101, Japan \\
$^{5}$Photocoding, 460-102 Iwakura-Nakamachi, Sakyo-ku, Kyoto 606-0025, Japan \\
$^{6}$Department of Astrophysics and Atmospheric Sciences, Faculty of Sciences, Kyoto Sangyo University, Motoyama, Kamigamo, \\ Kita-ku, Kyoto 603-8555, Japan \\
$^{7}$National Astronomical Observatory of Japan, 2-21-1 Osawa, Mitaka, Tokyo 181-8588, Japan}

\date{Accepted XXX. Received YYY; in original form ZZZ}
\pubyear{2017}
\begin{document}
\label{firstpage}
\pagerange{\pageref{firstpage}--\pageref{lastpage}}
\maketitle
\begin{abstract}
The effective temperature, one of the most fundamental atmospheric parameters of a star, can be estimated using various methods, and here we focus on the method using line-depth ratios (LDRs).
This method combines low- and high-excitation lines and makes use of relations between LDRs of these line pairs and the effective temperature.
It has an advantage, for example, of being minimally affected by interstellar reddening, which changes stellar colours.
We report 81 relations between LDRs and the effective temperature established with high-resolution, $\lambda /\Delta \lambda \sim 28,000$, spectra of nine G- to M-type giants in \textit{Y} and \textit{J} bands.
Our analysis gives the first comprehensive set of LDR relations for this wavelength range.
The combination of all these relations can be used to determine the effective temperatures of stars that have $3700<T_{\mathrm{eff}}<5400\ur{K}$ and $-0.5<\mathrm{[Fe/H]}<+0.3\ur{dex}$ to the precision of $\pm 10\ur{K}$ in the best cases.
\end{abstract}

\section{Introduction}
Stellar atmosphere is characterized mainly by effective temperature ($T_{\mathrm{eff}}$), surface gravity ($\log g$) and metallicity ([Fe/H]), of which $T_{\mathrm{eff}}$ has a particularly strong impact on the spectra and often is the first parameter to estimate during spectral analysis.
Among many methods of measuring $T_{\mathrm{eff}}$, we focus on the line-depth ratio (LDR) method in this paper.
In cool stars (around the solar temperature and below), the depths of low-excitation lines of neutral atoms are sensitive to $T_{\mathrm{eff}}$ while those of high-excitation lines are relatively insensitive \citep{2008oasp.book.....G}, and hence, their ratios are good temperature indicators \citep[and references therein]{1991PASP..103..439G,2015ApJ...812...64F}.
This method has a few advantages: LDRs are not affected by interstellar reddening, and atmospheric parameters other than $T_{\mathrm{eff}}$ are expected to have no, or at least relatively weak, effects on the LDRs.

\citet{1990ApJ...360..227S} used LDRs between \ion{C}{i} and \ion{Si}{i} lines around $1.1\,\text{\textmugreek m}$ to measure the temperatures and reddenings of Cepheids in a relative way.
This pioneering work, however, had a limited impact on later applications due to the small numbers of the standard stars and line pairs.
On the contrary, previous works on the LDR method mostly used visible spectral ranges\citep[e.g.][]{2003A&A...411..559K,2006MNRAS.371..879K,2007MNRAS.378..617K}.
With the development of infrared facilities in recent years, it has become possible to make more comprehensive studies using high-resolution infrared spectra.
Recently, \citet{2015ApJ...812...64F} established the LDR method for \textit{H}-band spectra, $1.50$--$1.65\,\text{\textmugreek m}$.
In this wavelength range, however, the small number of low-excitation lines and severe blending between molecular lines and metal lines make it difficult to find a large number of useful LDR pairs.
As a result, the precision of $T_{\mathrm{eff}}$ after taking an average of the values from available pairs is only $\sim 50\ur{K}$ for each star in the best case, while the LDR method in the visible range can achieve the precision of $\sim 5\ur{K}$ \citep{2006MNRAS.371..879K}.
Here we investigate \textit{Y}- and \textit{J}-band spectra to find the LDR--$T_{\mathrm{eff}}$ relations for the first time.
This wavelength range has a number of both low- and high-excitation lines, e.g. \citet{1999ApJS..124..527M}, and thus one can expect many useful LDR pairs to achieve a high precision.

\section{Observation}
The ten targets listed in \autoref{table:usedstar} were selected mainly from \textit{Gaia} FGK benchmark stars \citep{2014A&A...566A..98B,2014A&A...564A.133J,2015A&A...582A..49H} in addition to a few prototypes of red giants such as Arcturus.
All of our targets are giants with the luminosity class of III or II, and have the spectral types between early G to early M ($3700\ur{K}<T_{\mathrm{eff}}<5400\ur{K}$).
Their metallicities are between $-0.45$ and $+0.25\ur{dex}$ in [Fe/H] except Arcturus, $-0.52\ur{dex}$.

\begin{table*}
\caption{The target stars and their atmospheric parameters. Spectral types are from SIMBAD \citep{2000A&AS..143....9W}. The temperatures of \textmugreek \ Leo and Aldebaran were estimated in both references (1) and (2), and we use the weighted means. The last column gives the dates of our WINERED observations.}
\label{table:usedstar}
\begin{tabular}{lrlllll}\hline 
Object & HD & Sp. Type & $T_{\mathrm{eff}}\ \mathrm{[K]}$ & [Fe/H] [dex] & $\log g\ \mathrm{[dex]}$ & Obs. Date \\ \hline 
\textepsilon \ Leo & 84441 & G1II & $5398\pm 31^{\text{[2]}}$ & $-0.06\pm 0.04^{\text{[2]}}$ & $2.02\pm 0.08^{\text{[2]}}$ & 2014 Jan 23 \\
\textkappa \ Gem & 62345 & G8IIIa & $5029\pm 47^{\text{[2]}}$ & $-0.01\pm 0.05^{\text{[2]}}$ & $2.61\pm 0.12^{\text{[2]}}$ & 2013 Dec 8 \\
\textepsilon \ Vir & 113226 & G8III & $4983\pm 61^{\text{[1]}}$ & $+0.15\pm 0.16_{}^{\text{[1]}}$ & $2.77\pm 0.02^{\text{[1]}}$ & 2014 Jan 23 \\
Pollux & 62509 & K0III & $4858\pm 60^{\text{[1]}}$ & $+0.13\pm 0.16^{\text{[1]}}$ & $2.90\pm 0.08^{\text{[1]}}$ & 2013 Feb 28 \\
\textmugreek \ Leo & 85503 & K2IIIb & $4470\pm 40^{\text{[1,2]}}$ & $+0.25\pm 0.15^{\text{[1]}}$ & $2.51\pm 0.11^{\text{[1]}}$ & 2013 Feb 23 \\
Alphard & 81797 & K3II-III & $4171\pm 52^{\text{[2]}}$ & $+0.08\pm 0.07^{\text{[2]}}$ & $1.56\pm0.20^{\text{[2]}}$ & 2013 Nov 30 \\
Aldebaran & 29139 & K5III & $3882\pm 19^{\text{[1,2]}}$ & $-0.37\pm 0.17^{\text{[1]}}$ & $1.11\pm 0.19^{\text{[1]}}$ & 2013 Feb 24 \\
\textalpha \ Cet & 18884 & M1.5IIIa & $3796\pm 65^{\text{[1]}}$ & $-0.45\pm 0.47^{\text{[1]}}$ & $0.68\pm 0.23^{\text{[1]}}$ & 2013 Nov 30 \\
\textdelta \ Oph & 146051 & M0.5III & $3783\pm 20^{\text{[2]}}$ & $-0.03\pm 0.06^{\text{[2]}}$ & $1.45\pm 0.19^{\text{[2]}}$ & 2014 Jan 23 \\
Arcturus & 124897 & K0III & $4286\pm 35^{\text{[1]}}$ & $-0.52\pm 0.08^{\text{[1]}}$ & $1.64\pm 0.09^{\text{[1]}}$ & 2013 Feb 23 \\ \hline 
\end{tabular}

References: [1]~\citet{2015A&A...582A..49H} and [2]~\citet{2011A&A...531A.165P}.
\end{table*}

We observed these ten stars between 2013 February 23 and 2014 January 23 using WINERED attached to a Nasmyth platform of the $1.3\ur{m}$ Araki Telescope at Koyama Astronomical Observatory of Kyoto Sangyo University in Japan \citep{2016SPIE.9908E..5ZI}.
WINERED can collect spectra covering the wavelength range from $0.90$ to $1.35\,\text{\textmugreek m}$ (\textit{z$^{\prime }$}, \textit{Y} and \textit{J} bands) with the spectral resolution of $R\sim 28,000$ with one integration.
All of our targets are bright, $-2.3<J<2.0\ur{mag}$, and the total integration time with each target within the slit was between $12$ and $240\ur{sec}$ to achieve the signal-to-noise ratio (S/N) of $100$ or higher.
We describe the resultant S/N in \autoref{ssec:spectra_reduction}.
For every target star, we also observed a telluric standard star (an A0V star in most cases) for subtracting telluric absorption.

\section{Data Analysis}\label{sec:analysis}

\subsection{Spectra Reduction}\label{ssec:spectra_reduction}
\begin{figure}
\includegraphics[width=\columnwidth ]{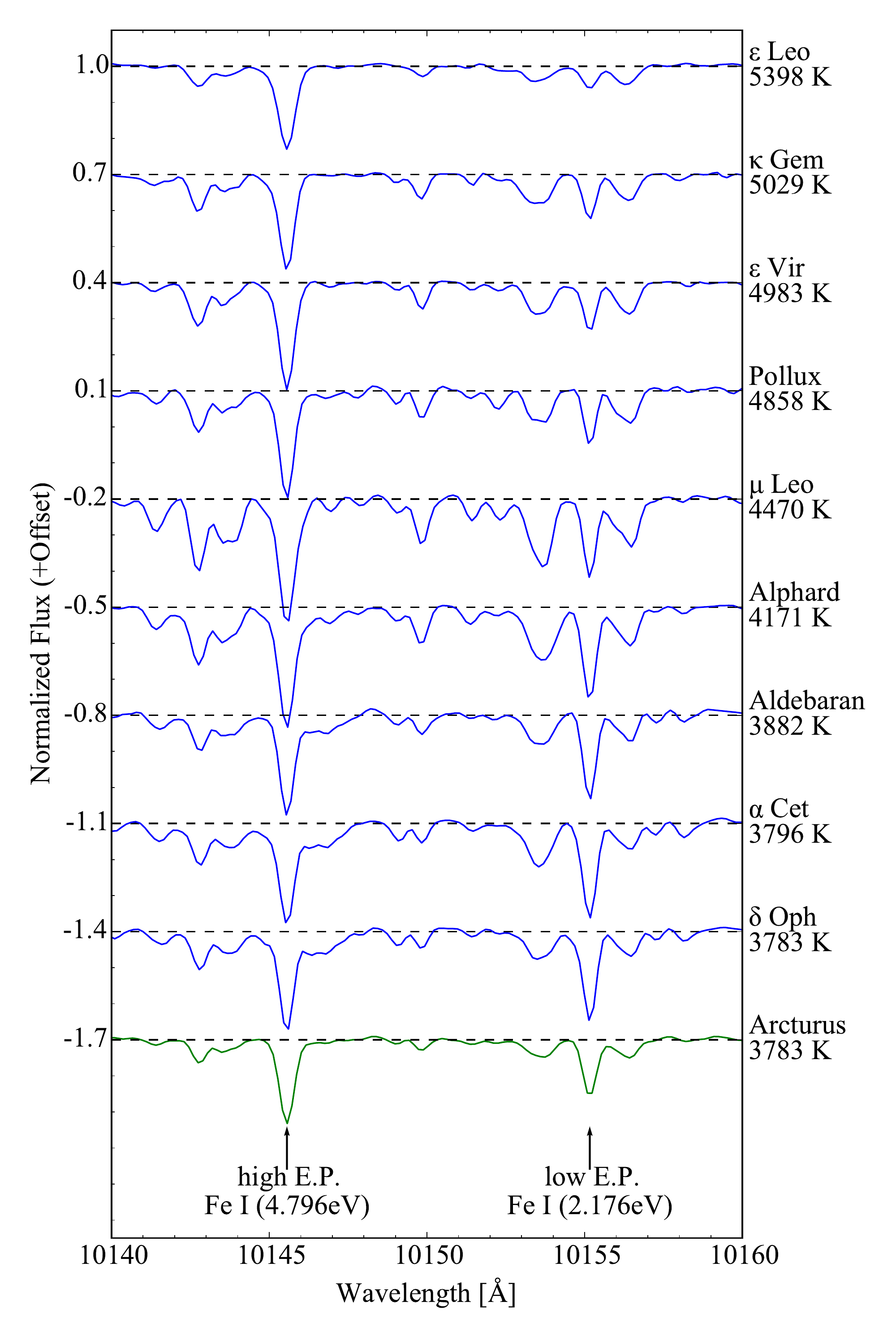}
\caption{A part of the spectra in the 55th order are drawn in the order of decreasing effective temperature (from top) except Arcturus in the bottom which were not included in establishing the LDR--$T_{\mathrm{eff}}$ relations. The two lines in the graph show a clear dependency on the temperature (the high-excitation line on the left and the low-excitation line on the right).}
\label{fig:compare_linedepth}
\end{figure}

Basic data reduction was performed automatically using the pipeline software prepared by the WINERED team based on \textsc{pyraf}\footnote{\textsc{pyraf} is a prodict of the Space Telescope Science Institute, which is operated by AURA for NASA.}.
The pipeline process includes the standard analysis steps for echelle spectra: bad pixel masking, sky subtraction, flat fielding, scattered light subtraction, transformation of each two-dimensional echelle image into images with the space and wavelength axes orthogonal to each other for individual orders, spectrum extraction, wavelength calibration and continuum normalisation.
In addition, it was necessary to correct for time-dependent wavelength shifts as follows.
We used ThAr lamp data as the initial wavelength calibration in the pipeline software, but we found that some spectra had wavelength offsets probably caused by varying ambient temperature\footnote{In late 2016, we made an instrumental upgrade to minimize such offsets.}.
In this work, such offsets are not critical because we are not interested in radial velocities, but the adjustment of the wavelength scale of individual frames was necessary to avoid the artificial broadening of the line spread function in a combined spectrum of each target.
The relative offsets between individual spectra were measured using almost isolated telluric absorption lines and were removed before combining the spectra.
Then, telluric absorption lines in the combined spectrum of each target were removed using a spectrum of a telluric-standard star except the 53rd to 55th orders, $1.01$ to $1.07\,\text{\textmugreek m}$, in which almost no significant telluric lines are present.
The wavelength scale of each telluric-corrected spectrum was converted to the rest frame of each star using its intrinsic absorption lines.
Our wavelengths are in the standard air scale.
Finally, we re-normalise the spectra using \textsc{pyraf} to adjust the continuum level to the unity.
In the following discussions, we consider 12 orders listed in \autoref{table:SN}.
We ignore other orders (42nd, 49th to 51st and 58th to 61st) which are around gaps between atmospheric windows.
In the analysis to find and calibrate the LDR--$T_{\mathrm{eff}}$ relations, we ignore Arcturus because, according to \citet{2015ApJ...812...64F}, its low metallicity causes offsets from the trends between the LDR and the effective temperature.
\autoref{table:SN} shows the S/N of each order of each star calculated from the variance of the normalised counts in continuum regions, as an indicator of spectral quality.
The spectral resolution was stable and quite constant for all of our spectra.
The FWHM values measured with most stellar lines correspond to $\lambda /\Delta \lambda $ between $26000$ and $30000$.
\autoref{fig:compare_linedepth} shows a small range of the obtained spectra.
This range has a line pair which shows a clear temperature dependency as we discuss below.

\begin{table*}
\caption{S/N of the reduced spectra in each echelle order (52--57th in \textit{Y} band and 43--48th in \textit{J} band).}
\label{table:SN}
\begin{tabular}{lrrrrrrrrrrrr} \hline 
Object & 57 & 56 & 55 & 54 & 53 & 52 & 48 & 47 & 46 & 45 & 44 & 43 \\ \hline 
\textepsilon \ Leo & $321$ & $293$ & $581$ & $266$ & $332$ & $207$ & $173$ & $178$ & $194$ & $224$ & $198$ & $178$ \\
\textkappa \ Gem & $102$ & $154$ & $385$ & $406$ & $248$ & $192$ & $141$ & $269$ & $170$ & $359$ & $245$ & $233$ \\
\textepsilon \ Vir & $99$ & $86$ & $289$ & $532$ & $290$ & $172$ & $111$ & $200$ & $204$ & $329$ & $220$ & $225$ \\
Pollux & $134$ & $117$ & $150$ & $160$ & $287$ & $126$ & $121$ & $120$ & $143$ & $261$ & $152$ & $187$ \\
\textmugreek \ Leo & $87$ & $82$ & $175$ & $353$ & $226$ & $126$ & $102$ & $118$ & $99$ & $166$ & $119$ & $144$ \\
Alphard & $128$ & $119$ & $263$ & $254$ & $291$ & $155$ & $103$ & $89$ & $124$ & $223$ & $97$ & $146$ \\
Aldebaran & $49$ & $153$ & $173$ & $176$ & $141$ & $192$ & $185$ & $137$ & $174$ & $312$ & $166$ & $181$ \\
\textalpha \ Cet & $125$ & $134$ & $143$ & $226$ & $290$ & $229$ & $166$ & $165$ & $151$ & $342$ & $241$ & $171$ \\
\textdelta \ Oph & $195$ & $140$ & $142$ & $301$ & $368$ & $186$ & $103$ & $177$ & $165$ & $405$ & $186$ & $293$ \\
Arcturus & $118$ & $192$ & $246$ & $248$ & $264$ & $191$ & $128$ & $99$ & $188$ & $193$ & $158$ & $278$ \\ \hline 
\end{tabular}
\end{table*}

\subsection{Line Selection and Measurement of Line Depth}
As the candidates of atomic lines to be used for LDRs, we consider the lines identified in our Arcturus spectrum.
Here we outline its line identification, but details of the analysis and the entire line list will be published in a forthcoming paper (Ikeda et al. in preparation).
Firstly, we produced a synthesized spectrum of Arcturus, whose parameters are assumed to be $T_{\mathrm{eff}}=4275\ur{K}$, $\log g=1.7$, and $\text{[Fe/H]}=-0.4\ur{dex}$ \citep{2013ApJ...765...16S} using the ATLAS9 atmospheric models and codes given by \citet{1993KurCD..13.....K}.
As the input list of atomic lines for the synthetic spectrum, Vienna Atomic Line Database \citep[VALD3;][]{2015PhyS...90e4005R} were used.
Secondly, the observed and synthetic spectra were closely compared with each other to identify lines whose depth is larger than $\sim 1\%$.
Although the S/N of the Arcturus spectrum is higher than $200$ in all wavelengths, CN molecular lines often make the identification of shallow lines less certain.
In such cases, to confirm the contribution of weak atomic lines around molecular lines, we further compared the observed spectrum with a synthetic spectrum but with only CN lines included.
Thus we obtained approximately 800 atomic lines.

Among the lines identified in Arcturus, we here include relatively isolated neutral lines of Fe, Ti, Si, Cr, Ca, Ni, Mg, Na, Co, Al, Mn and K in our analysis for LDRs. Ion lines and lines of C, N and O were not included \citep{2006MNRAS.371..879K}.
Moreover, we excluded lines by the following criteria: deep lines which appear deeper than $\sim 0.5$ in Arcturus, and lines blended with molecular CN lines or other atomic lines which have different elements and/or excitation potentials.
These criteria left 125 and 99 lines in \textit{Y} and \textit{J} bands, respectively.
We refer to VALD3 for parameters of the lines such as excitation potential and oscillator strength.

For the 125 and 99 lines we selected, we used a quadratic function to fit three (or four) pixels near the bottom of each line rather than fitting the entire line profile, e.g., by the Gaussian or the Voigt function \citep{2000AN....321..277S}.
We define $d_{i}^{(n)}$ as a line depth from the continuum level to the bottom of the fitted function, where $i$ indicates an ID number of the line and $(n)$ indicates an ID number of a star.
Weak lines with $d_{i}^{(n)}<0.02$ are ignored in the following analysis.

\subsection{Line Pair Selection}
We calculated an LDR, $r_{j}^{(n)}=d_{i}^{(n)}/d_{i'}^{(n)}$, for any line pair, $j$, of approximately 1500 pairs whose lines were both detected in more than four stars and have excitation potentials separated by more than $1\ur{eV}$.
We treated each order independently, i.e. did not combine lines in different orders, and divided the depth of the low-excitation line, $d_{i}$, by that of the high-excitation line, $d_{i'}$.
For each line pair, $j$, we plotted the effective temperatures $T_{\mathrm{eff}}$ against the common logarithms of LDRs, $\log r_{j}$, and determined the regression line, $T_{\mathrm{eff}}=a_{j}\log r_{j}+b_{j}$, using the Weighted Total Least Squares method \citep[see][for a review]{markovsky2007overview}.
The weight of a line pair, $j$, of a star $(n)$ was calculated by $w_{j}^{(n)}=\left[(\sigma _{y}^{(n)})^{2}+{a_{j}}^{2}(\sigma _{x,j}^{(n)})^{2}\right]^{-1}$, where $\sigma _{x,j}^{(n)}$ and $\sigma _{y}^{(n)}$ indicate the standard error of $\log r_{j}^{(n)}$ calculated by the S/N and that of literature $T_{\mathrm{eff}}^{(n)}$ for each star, respectively.
The dispersion around each regression line is defined as ${\sigma _{j}}^{2}$.
The effective temperature and its error of each star based on each relation are given as $T_{j}^{(n)}\pm \Delta T_{j}^{(n)}$, of which the error was determined by variance--covariance matrix of the coefficients $(a_{j},b_{j})$ of the regression line and the error of $\log r_{j}^{(n)}$.
In the subsequent analysis, about 900 line pairs which have $\sigma _{j}>150\ur{K}$ were not included.
Moreover, 21 line pairs which have $a_{j}>0$ were excluded.
A large fraction of the rejected pairs includes a weak iron line at $10350.77\,\text{\AA }$ ($6.145\ur{eV}$) in \textit{Y} band.
This line may be blended by other line(s), which causes the unexpected dependency of LDR on the effective temperature, though the presence of such blended lines is not clear in available line lists.
The rest of the line pairs with $a_{j}>0$ have \ion{Ti}{i} lines for both low- and high-excitation potentials \footnote{We found relatively tight LDR--$T_{\mathrm{eff}}$ relations with the unexpected slopes for six pairs of \ion{Ti}{i} lines (for each pair, the first is the low-excitation line): $10496.11\,\text{\AA }$--$10551.76\,\text{\AA }$, $10607.72\,\text{\AA }$--$10551.76\,\text{\AA }$, $12671.10\,\text{\AA }$--$12744.91\,\text{\AA }$, $12821.67\,\text{\AA }$--$12744.91\,\text{\AA }$, $12831.44\,\text{\AA }$--$12744.91\,\text{\AA }$ and $12847.03\,\text{\AA }$--$12744.91\,\text{\AA }$.}.
LDR--$T_{\mathrm{eff}}$ relations for the majority of \ion{Ti}{i}--\ion{Ti}{i} pairs are not tight enough to be selected, and their slopes can be negative, positive or not well determined.
The reason of the opposite slopes for the small number of \ion{Ti}{i}--\ion{Ti}{i} pairs is unclear.

There is a large number of possible combinations of line pairs, and we selected a set of pairs as follows.
One condition is that each line is used only once, i.e. not included in more than one line pairs.
This ensures that statistical errors in $r_{j}^{(n)}$ values in different LDR--$T_{\mathrm{eff}}$ relation are independent of each other and makes it easy to calculate the statistical errors of combined effective temperatures ($T_{\mathrm{LDR}}$ in \autoref{sec:Result}).
In fact, the final set of the relations we discuss in \autoref{sec:Result} makes use of 162 lines, which is about $70\%$ of the selected lines.
This condition for the line pair selection has no large impact on the statistical error, which can be better only by $\sim 15\%$ even if we use all the pre-selected pairs.

The basic idea of the process is to select line pairs which meet the following conditions as much as possible: (i)~high precision in reproducing the effective temperatures of our sample stars, and (ii)~small difference in wavelength between two lines of each line pair.
With the latter condition fulfilled, the possibility for other instruments to detect both lines of a line pair in the same echelle order gets high and the error of the LDR introduced in the continuum normalisation is expected to be smaller.

First, we consider the undirected graph whose nodes correspond to absorption lines and edges connecting the nodes correspond to line pairs.
Note that our analysis treats each order independently.
The goal is to find the optimal matching, $M$, that meets the above conditions for each order.
We consider maximum matchings, the number of whose edges is as large as possible but no node is connected by more than one edges.
In an ideal case, the size of a maximum matching, $\abs{M}$, corresponds to half the number of all the nodes in the original undirected graph, but this is not true in our case because many edges were rejected owing to, for example, large scatters around the corresponding LDR--$T_{\mathrm{eff}}$ relations.
For a maximum matching, $M_{k}$, of this undirected graph, each line pair of $M_{k}$ was named $j_{k,1},\cdots ,j_{k,m},\cdots ,j_{k,\abs{M_{k}}}$.
For a given matching, each star's effective temperature was redetermined as a weighted mean of $T_{j_{k,m}}^{(n)}\pm \Delta T_{j_{k,m}}^{(n)}$ and was named $T_{M_{k}}^{(n)}$.
The statistical error in $T_{M_{k}}^{(n)}$ was estimated by a weighted standard deviation of $T_{j_{k,m}}^{(n)}$.
Then, let $\Delta \lambda _{k,m}$ be the difference in wavelength between two lines of a line pair, $j_{k,m}$, and we consider the evaluation function $E(M_{k};e)$ as 
\begin{align}\label{eq:evalfunc}
E(M_{k};e)&=\sqrt{\frac{1}{N}\sum _{n=1}^{N}\left(T_{M_{k}}^{(n)}-T_{\mathrm{eff}}^{(n)}\right)^{2}}+e\sqrt{\frac{1}{\abs{M_{k}}}\sum _{m=1}^{\abs{M_{k}}}\left(\Delta \lambda _{k,m}\right)^{2}} \\
&=E_{T}(M_{k})+eE_{\lambda }(M_{k})\text{.}
\end{align}
$E_{T}(M_{k})$ represents the size of the error in redetermining the effective temperatures of the nine stars, without Arcturus included, for a given matching $M_{k}$, while $E_{\lambda }(M_{k})$ represents the wavelength difference of the line pairs and works as a penalty term.
For a given $e$ value, there are different allowed combinations of line pairs which form different maximum matchings, and we select the one which gives the least $E(M_{k};e)$ as the optimal matching, $M_{k(e)}$.
The coefficient $e$ will be determined, in the next section, by considering how $E_{T}(M_{k(e)};e)$ and $E_{\lambda }(M_{k(e)};e)$ depend on it.

\section{Result}\label{sec:Result}
\begin{figure*}
\includegraphics[width=2\columnwidth ]{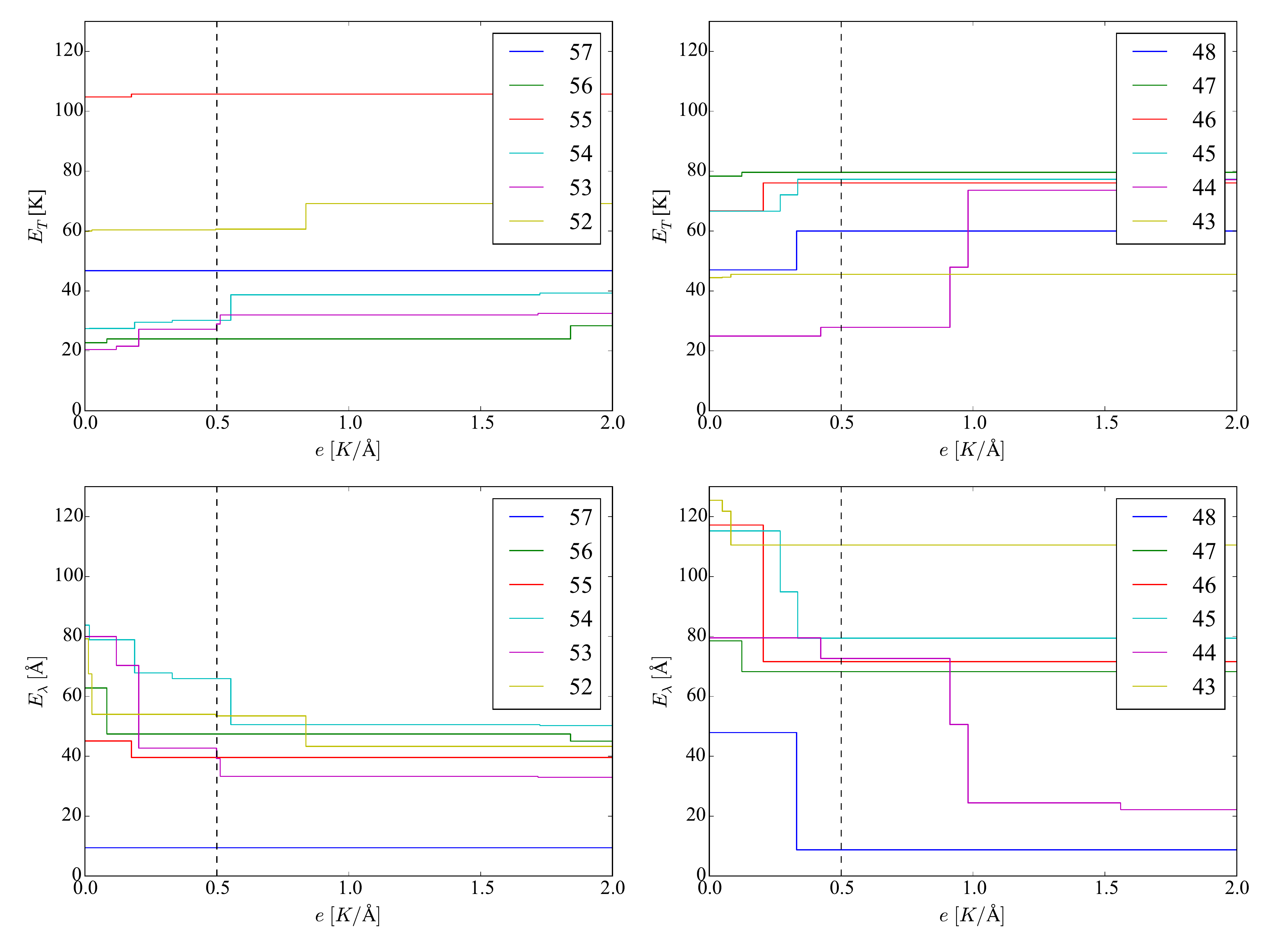}
\caption{The dependence of $E_{T}(M_{k(e)})$ and $E_{\lambda }(M_{k(e)})$ on the parameter, $e$, in \autoref{eq:evalfunc}. The evaluation functions are defined in the text. Variations of $E_{T}(M_{k(e)})$ (top row) and $E_{\lambda }(M_{k(e)})$ (bottom row) are presented for a group of echelle orders in each band (52nd to 57th in \textit{Y} band on the left and 43rd to 48th in \textit{J} band on the right). Each solid line represents the evaluation function and the vertical dashed line represents the value $e=0.5\ur{K/\text{\AA }}$ which we used for the final solutions. Legends show which colour corresponds to which echelle order. The colour version of this figure is available in the online journal.}
\label{fig:evalfunc}
\end{figure*}

We applied the procedure described above to our 224 nodes (i.e. lines) and 603 edges (i.e. line pairs).
With changing the coefficient $e$ from $0$ to $2\ur{K/\text{\AA }}$, we searched for the optimized matching, $M_{k(e)}$, which gives the smallest $E$ at each $e$ value and observed how $e$ affects the solutions.
\autoref{fig:evalfunc} plots the values of $E_{T}$ and $E_{\lambda }$ for $M_{k(e)}$ with varying $e$.
By increasing $e$, the weight of $E_{\lambda }$ increases relative to $E_{T}$ in the evaluation function, and then the optimal matching and the values of $E_{T}$ and $E_{\lambda }$ change.
At around $e=0.5\ur{K/\text{\AA }}$, $E_{T}$ is only slightly larger compared to the case with $e=0\ur{K/\text{\AA }}$ for most orders (the exception being the order 48) which optimizes the precision in the redetemined temperature by ignoring the difference in the wavelength, while $E_{\lambda }$ improves to some extent by changing $e$ from $0$ to $0.5\ur{K/\text{\AA }}$.
The number of line pairs in the 48th order is only 4, and with $e=0.5\ur{K/\text{\AA }}$ the wavelength separations of the individual pairs get significantly small with only a small increase in $E_{T}$.
We selected $e=0.5\ur{K/\text{\AA }}$, and \autoref{table:evallambda} lists the resultant values of the evaluation functions and other parameters in our analysis.

\begin{table*}
\caption{Parameters of the selected line pairs in individual orders (52--57th in \textit{Y} and 43--48th in \textit{J}); the wavelength range $\lambda _{\mathrm{min}}<\lambda <\lambda _{\mathrm{max}}$ from the WINERED web site (http://merlot.kyoto-su.ac.jp/LIH/WINERED/), the number of lines $N_{\mathrm{line}}$ considered to make pairs, the number of line pairs $N_{\mathrm{pair}}$ which we included in the final solutions and the values of the evaluation functions. In addition, the last column gives the same parameters but for all the orders combined.}
\label{table:evallambda}
\begin{tabular}{lrrrrrrrrrrrrr} \hline 
Order & 57 & 56 & 55 & 54 & 53 & 52 & 48 & 47 & 46 & 45 & 44 & 43 & Combined \\ \hline 
$\lambda _{\mathrm{min}}\ $[\textmugreek m] & $0.976$ & $0.992$ & $1.010$ & $1.028$ & $1.048$ & $1.068$ & $1.156$ & $1.180$ & $1.205$ & $1.232$ & $1.260$ & $1.290$ & $0.976$ \\
$\lambda _{\mathrm{max}}\ $[\textmugreek m] & $0.992$ & $1.010$ & $1.028$ & $1.048$ & $1.068$ & $1.089$ & $1.180$ & $1.205$ & $1.232$ & $1.260$ & $1.290$ & $1.319$ & $1.319$ \\
$N_{\mathrm{line}}$ & 4 & 23 & 19 & 30 & 17 & 32 & 13 & 11 & 19 & 14 & 16 & 26 & 224 \\
$N_{\mathrm{pair}}$ & 1 & 10 & 4 & 10 & 8 & 14 & 3 & 3 & 4 & 5 & 7 & 12 & 81 \\
$E_{T}(e=0.5)\ \mathrm{[K]}$ & $47$ & $24$ & $106$ & $30$ & $29$ & $61$ & $60$ & $80$ & $76$ & $77$ & $28$ & $46$ & $28.6$ \\
$E_{\lambda }(e=0.5)\ $[\AA ] & $9$ & $47$ & $40$ & $66$ & $39$ & $53$ & $9$ & $68$ & $72$ & $79$ & $73$ & $111$ & $68.0$ \\
$E(e=0.5)\ $[K] & $51$ & $48$ & $126$ & $63$ & $49$ & $87$ & $64$ & $114$ & $112$ & $117$ & $64$ & $101$ & $62.6$ \\ \hline 
\end{tabular}
\end{table*}

\begin{figure}
\includegraphics[width=\columnwidth ]{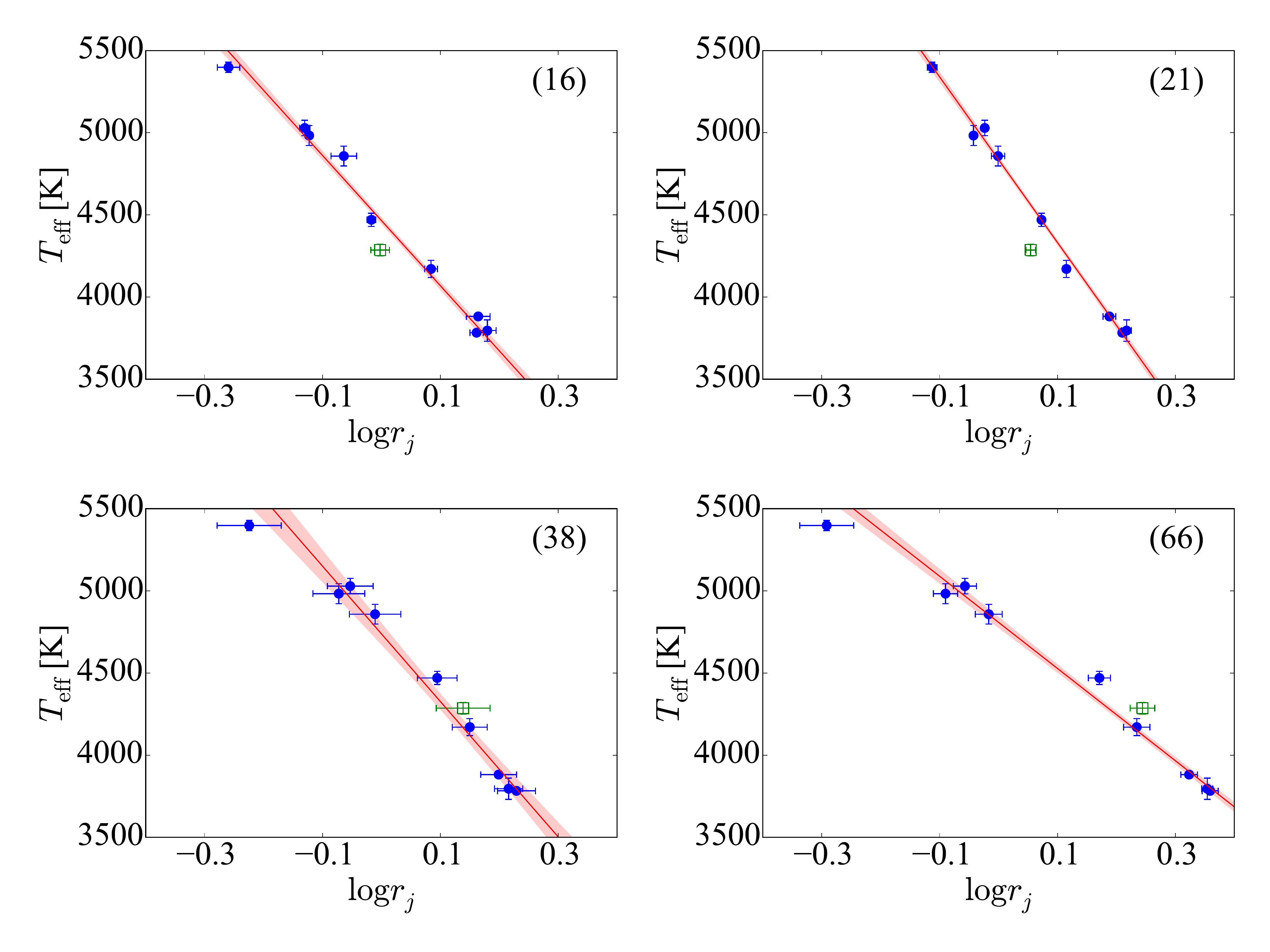}
\caption{Examples of the LDR--$T_{\mathrm{eff}}$ relations. The ID of the line pair is given in each panel. Arcturus indicated by a green open square in each panel was not used for the relation which was obtained with the other objects indicated by blue filled circles. Plots for all the 81 relations are available as online material --- see Surpporting Information.}
\label{fig:LDR_some}
\end{figure}

\begin{figure*}
\includegraphics[width=2\columnwidth ]{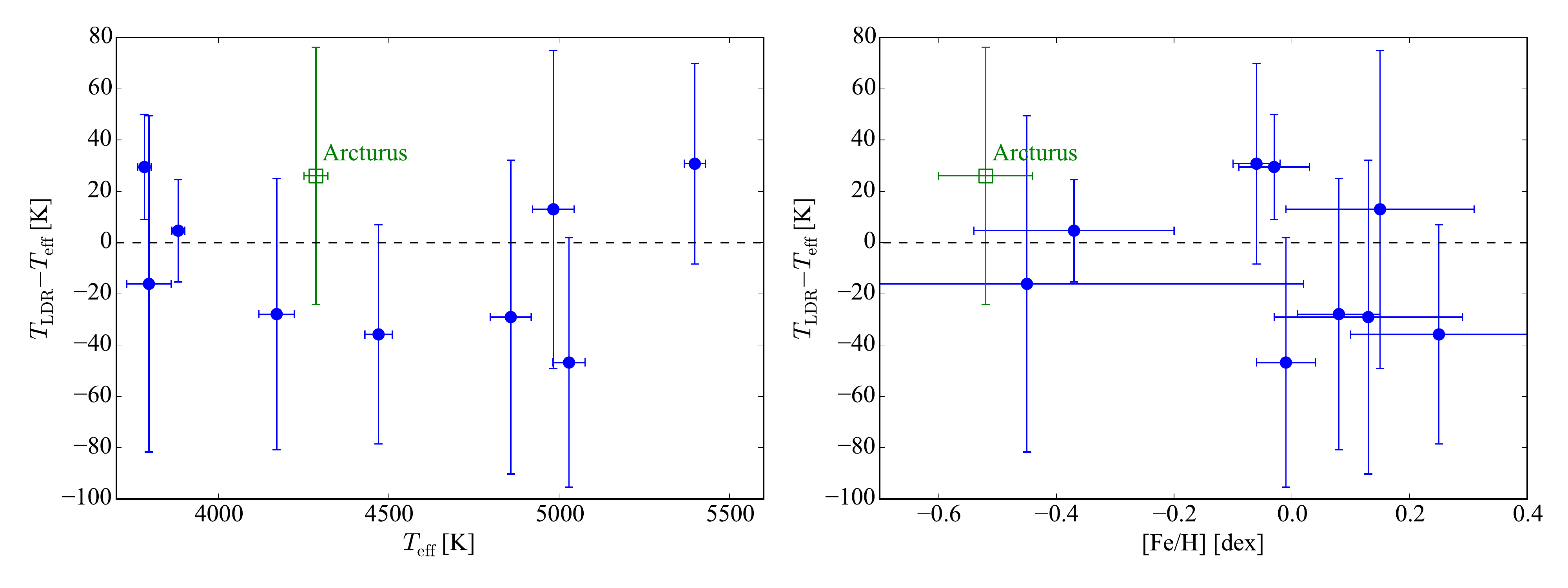}
\caption{The differences between the redetermined effective temperatures $T_{\mathrm{LDR}}$ by our LDR method and the literature temperatures $T_{\mathrm{eff}}$ are plotted against $T_{\mathrm{eff}}$ (left panel) and [Fe/H] (right panel). Symbols are same as in \autoref{fig:LDR_some}. The size of the vertical error bars is dominated by the error in the literature temperature except for high-temperature stars (\textepsilon \ Leo and \textkappa \ Gem) and Arcturus.}
\label{fig:reesttemp}
\end{figure*}

We obtained 47 LDR--$T_{\mathrm{eff}}$ relations in \textit{Y} band and 34 in \textit{J} band with $e=0.5\ur{K/\text{\AA }}$.
\autoref{table:resultY} and \ref{table:resultJ} list the line pairs and the relations, and \autoref{fig:LDR_some} shows some examples of the relations.
We redetermined the effective temperatures using our final relations ($T_{\mathrm{LDR}}$ in \autoref{table:reesttemp_nottre}) and compared them with $T_{\mathrm{eff}}$ in the literature in \autoref{fig:reesttemp}.
The difference, $T_{\mathrm{LDR}}-T_{\mathrm{eff}}$, has no clear dependency on $T_{\mathrm{eff}}$ or [Fe/H], suggesting that this LDR method is effective across the parameter range covered by the nine calibrating stars, i.e. $3700<T_{\mathrm{eff}}<5400\ur{K}$ and $-0.5<\mathrm{[Fe/H]}<+0.3\ur{dex}$.
In order to evaluate the power of the LDR method considering all the orders available, we consider the evaluation function of \autoref{eq:evalfunc} but including all the orders and the line pairs: i.e. we use $T_{\mathrm{LDR}}^{(n)}$ instead of the temperatures for an individual order, $T_{M_{k(e)}}^{(n)}$, and consider the wavelength differences of all the line pairs.
Thus calculated evaluation functions are $E_{T}^{\mathrm{all}}=28.6\ur{K}$ and $E_{\lambda }^{\mathrm{all}}=68.0\ur{\text{\AA }}$.
These values are, for example, smaller than those\footnote{\citet{2015ApJ...812...64F} used no evaluation function like ours, but we used their published data to calculate the values according to \autoref{eq:evalfunc}.} for the result by \citet{2015ApJ...812...64F}, $E_{T}^{\mathrm{IRCS}}=85.5\ur{K}$ and $E_{\lambda }^{\mathrm{IRCS}}=295.6\ur{\text{\AA }}$, who used nine line pairs in \textit{H} band.

It is worthwhile to discuss $T_{\mathrm{LDR}}$ for Arcturus, which was not included in the calibration.
\citet{2015ApJ...812...64F} found that Arcturus tends to show offsets from LDR--$T_{\mathrm{eff}}$ relations of other stars.
They suggested that this is owing to the low metallicity and/or non-solar abundance ratios of Arcturus, $\mathrm{[Fe/H]}=-0.52$, $\mathrm{[Ti/H]}=-0.313$, $\mathrm{[Si/H]}=-0.252$ by \citet{2015A&A...582A..81J}.
The result in \autoref{table:reesttemp_nottre} appears to show that the temperature from our method ($T_{\mathrm{LDR}}=4312\pm 36\ur{K}$) is consistent with the literature $T_{\mathrm{eff}}$, but a closer look reveals the peculiarity of Arcturus's $T_{\mathrm{LDR}}$.
If we use only 22 line pairs composed of \ion{Ti}{i} (low-excitation) and \ion{Fe}{i} (high-excitation), we obtain the temperature of $4120\pm 60\ur{K}$.
If we use only 14 line pairs of \ion{Si}{i} (high-excitation) and \ion{Fe}{i} (low-excitation), in contrast, the temperature is estimated at $4485\pm 35\ur{K}$.
These deviations can be understood as the effect of non-solar [Ti/Fe] and [Si/Fe] values.
These offsets of the two major groups of line pairs cancelled out each other, and the average $T_{\mathrm{LDR}}$ became close to the literature temperature , but giving a relatively large error, in the case of Arcturus.
Such deviations are not observed in other objects in this study.
Follow-up spectroscopic data for a larger sample of stars with known temperatures and abundances would be useful to further discuss the abundance dependences of the LDR--$T_{\mathrm{eff}}$ relations.

\begin{table*}
\caption{List of low- and high-excitation lines and the LDR--$T_{\mathrm{eff}}$ relations in \textit{Y} band. $a$ and $b$ represent the coefficients of $T_{\mathrm{eff}}=a\log r+b$, $N$ represents the number of stars used in fitting and $\sigma $ represents the dispersion of the relation. If more than one lines are listed at the same wavelength in the original line catalogue, information of the line with the largest $\log gf$ is adopted here. The flag in the last column indicates the line pairs that are used in the analysis without the telluric correction in \autoref{sec:discussion}.}

\label{table:resultY}
\begin{tabular}{ccrcrrcrrrcrr} \hline 
 & & \multicolumn{3}{c}{Low-excitation Line} & \multicolumn{3}{c}{High-excitation Line} & \multicolumn{4}{c}{LDR--$T_{\mathrm{eff}}$ Relation} & \\
ID & Order & \multicolumn{1}{c}{$\lambda $ [\AA ]} & El. & \multicolumn{1}{c}{E.P. [eV]} & \multicolumn{1}{c}{$\lambda $ [\AA ]} & El. & \multicolumn{1}{c}{E.P. [eV]} & \multicolumn{1}{c}{$a$ [K]} & \multicolumn{1}{c}{$b$ [K]} & $N$ & \multicolumn{1}{c}{$\sigma \ \mathrm{[K]}$} & flag \\ \hline 
(1) & 57 & $9879.583$ & $\ion{Ti}{i}$ & $1.873$ & $9889.035$ & $\ion{Fe}{i}$ & $5.033$ & $-1462$ & $3543$ & $5$ & $43$ & \\

(2) & 56 & $9927.351$ & \ion{Ti}{i} & $1.879$ & $9944.207$ & \ion{Fe}{i} & $5.012$ & $-2390$ & $4837$ & $8$ & $112$ & \\
(3) & 56 & $9946.320$ & \ion{Cr}{i} & $3.556$ & $9953.471$ & \ion{Fe}{i} & $5.446$ & $-2635$ & $4223$ & $6$ & $125$ & 1 \\
(4) & 56 & $10003.09$ & \ion{Ti}{i} & $2.160$ & $9961.256$ & \ion{Na}{i} & $3.617$ & $-2645$ & $4339$ & $6$ & $86$ & 1 \\
(5) & 56 & $10081.39$ & \ion{Fe}{i} & $2.424$ & $9980.463$ & \ion{Fe}{i} & $5.033$ & $-2687$ & $4586$ & $9$ & $74$ & \\
(6) & 56 & $10005.66$ & \ion{Ti}{i} & $1.067$ & $9987.869$ & \ion{Fe}{i} & $2.176$ & $-961$ & $3955$ & $5$ & $27$ & \\
(7) & 56 & $9997.959$ & \ion{Ti}{i} & $1.873$ & $10077.17$ & \ion{Fe}{i} & $2.990$ & $-2297$ & $4848$ & $8$ & $121$ & \\
(8) & 56 & $10011.74$ & \ion{Ti}{i} & $2.154$ & $10041.47$ & \ion{Fe}{i} & $5.012$ & $-2477$ & $5189$ & $9$ & $87$ & \\
(9) & 56 & $10034.49$ & \ion{Ti}{i} & $1.460$ & $10086.24$ & \ion{Fe}{i} & $2.949$ & $-2433$ & $4953$ & $8$ & $42$ & \\
(10) & 56 & $10059.90$ & \ion{Ti}{i} & $1.430$ & $10068.33$ & \ion{Si}{i} & $6.099$ & $-1088$ & $4837$ & $6$ & $99$ & \\
(11) & 56 & $10066.51$ & \ion{Ti}{i} & $2.160$ & $10065.05$ & \ion{Fe}{i} & $4.835$ & $-1801$ & $3631$ & $9$ & $79$ & 1 \\

(12) & 55 & $10155.16$ & \ion{Fe}{i} & $2.176$ & $10216.31$ & \ion{Fe}{i} & $4.733$ & $-3116$ & $3758$ & $9$ & $148$ & \\
(13) & 55 & $10167.47$ & \ion{Fe}{i} & $2.198$ & $10193.22$ & \ion{Ni}{i} & $4.089$ & $-4235$ & $4587$ & $9$ & $143$ & \\
(14) & 55 & $10189.15$ & \ion{Ti}{i} & $1.460$ & $10195.11$ & \ion{Fe}{i} & $2.728$ & $-1654$ & $3522$ & $6$ & $78$ & \\
(15) & 55 & $10273.68$ & \ion{Ca}{i} & $4.535$ & $10230.80$ & \ion{Fe}{i} & $6.119$ & $-2179$ & $4584$ & $9$ & $102$ & \\

(16) & 54 & $10423.74$ & \ion{Fe}{i} & $3.071$ & $10288.94$ & \ion{Si}{i} & $4.920$ & $-3964$ & $4465$ & $9$ & $69$ & 1 \\
(17) & 54 & $10333.18$ & \ion{Fe}{i} & $4.593$ & $10313.20$ & \ion{Si}{i} & $6.399$ & $-2830$ & $5109$ & $5$ & $68$ & 1 \\
(18) & 54 & $10321.06$ & \ion{Ni}{i} & $5.525$ & $10414.91$ & \ion{Si}{i} & $6.619$ & $-3392$ & $4482$ & $5$ & $132$ & 1 \\
(19) & 54 & $10332.33$ & \ion{Fe}{i} & $3.635$ & $10353.81$ & \ion{Fe}{i} & $5.393$ & $-4499$ & $4779$ & $9$ & $96$ & 1 \\
(20) & 54 & $10396.80$ & \ion{Ti}{i} & $0.848$ & $10340.89$ & \ion{Fe}{i} & $2.198$ & $-4433$ & $5016$ & $9$ & $110$ & 1 \\
(21) & 54 & $10343.82$ & \ion{Ca}{i} & $2.933$ & $10371.26$ & \ion{Si}{i} & $4.930$ & $-5040$ & $4836$ & $9$ & $49$ & 1 \\
(22) & 54 & $10423.03$ & \ion{Fe}{i} & $2.692$ & $10347.97$ & \ion{Fe}{i} & $5.393$ & $-3912$ & $5346$ & $9$ & $127$ & 1 \\
(23) & 54 & $10382.31$ & \ion{Co}{i} & $2.871$ & $10388.75$ & \ion{Fe}{i} & $5.446$ & $-2256$ & $4454$ & $7$ & $97$ & 1 \\
(24) & 54 & $10395.80$ & \ion{Fe}{i} & $2.176$ & $10469.65$ & \ion{Fe}{i} & $3.884$ & $-8570$ & $4524$ & $9$ & $146$ & 1 \\
(25) & 54 & $10460.05$ & \ion{Ti}{i} & $2.256$ & $10435.35$ & \ion{Fe}{i} & $4.733$ & $-2058$ & $4693$ & $8$ & $117$ & 1 \\

(26) & 53 & $10486.25$ & \ion{Cr}{i} & $3.011$ & $10530.51$ & \ion{Ni}{i} & $4.105$ & $-3839$ & $5192$ & $9$ & $70$ & 1 \\
(27) & 53 & $10496.11$ & \ion{Ti}{i} & $0.836$ & $10535.71$ & \ion{Fe}{i} & $6.206$ & $-1832$ & $5620$ & $9$ & $107$ & 1 \\
(28) & 53 & $10510.01$ & \ion{Cr}{i} & $3.013$ & $10555.65$ & \ion{Fe}{i} & $5.446$ & $-2015$ & $5081$ & $9$ & $92$ & 1 \\
(29) & 53 & $10551.76$ & \ion{Ti}{i} & $1.887$ & $10532.24$ & \ion{Fe}{i} & $3.929$ & $-1475$ & $3325$ & $5$ & $39$ & 1 \\
(30) & 53 & $10552.97$ & \ion{Ti}{i} & $2.249$ & $10582.16$ & \ion{Si}{i} & $6.223$ & $-936$ & $4393$ & $7$ & $44$ & 1 \\
(31) & 53 & $10577.14$ & \ion{Fe}{i} & $3.301$ & $10611.69$ & \ion{Fe}{i} & $6.169$ & $-4007$ & $4805$ & $9$ & $106$ & 1 \\
(32) & 53 & $10607.72$ & \ion{Ti}{i} & $0.848$ & $10672.14$ & \ion{Cr}{i} & $3.013$ & $-2069$ & $4236$ & $8$ & $91$ & 1 \\
(33) & 53 & $10616.72$ & \ion{Fe}{i} & $3.267$ & $10627.65$ & \ion{Si}{i} & $5.863$ & $-2750$ & $3799$ & $9$ & $106$ & 1 \\

(34) & 52 & $10677.05$ & \ion{Ti}{i} & $0.836$ & $10721.66$ & \ion{Fe}{i} & $5.507$ & $-1297$ & $5420$ & $7$ & $117$ & \\
(35) & 52 & $10692.73$ & \ion{Fe}{i} & $3.071$ & $10689.72$ & \ion{Si}{i} & $5.954$ & $-2088$ & $2691$ & $6$ & $113$ & \\
(36) & 52 & $10753.00$ & \ion{Fe}{i} & $3.960$ & $10694.25$ & \ion{Si}{i} & $5.964$ & $-3586$ & $3739$ & $9$ & $103$ & \\
(37) & 52 & $10746.45$ & \ion{Na}{i} & $3.191$ & $10717.81$ & \ion{Fe}{i} & $5.539$ & $-4899$ & $6654$ & $7$ & $133$ & \\
(38) & 52 & $10725.19$ & \ion{Fe}{i} & $3.640$ & $10849.47$ & \ion{Fe}{i} & $5.539$ & $-4120$ & $4739$ & $9$ & $97$ & \\
(39) & 52 & $10726.39$ & \ion{Ti}{i} & $0.813$ & $10785.39$ & \ion{Fe}{i} & $5.621$ & $-1685$ & $5490$ & $8$ & $144$ & \\
(40) & 52 & $10783.05$ & \ion{Fe}{i} & $3.111$ & $10727.41$ & \ion{Si}{i} & $5.984$ & $-3629$ & $3936$ & $9$ & $69$ & \\
(41) & 52 & $10732.86$ & \ion{Ti}{i} & $0.826$ & $10762.26$ & \ion{Ni}{i} & $4.154$ & $-1681$ & $5364$ & $8$ & $70$ & \\
(42) & 52 & $10742.55$ & \ion{Fe}{i} & $3.642$ & $10749.38$ & \ion{Si}{i} & $4.930$ & $-4284$ & $-120$ & $5$ & $145$ & \\
(43) & 52 & $10774.87$ & \ion{Ti}{i} & $0.818$ & $10838.97$ & \ion{Ca}{i} & $4.878$ & $-1628$ & $4756$ & $8$ & $74$ & \\
(44) & 52 & $10780.69$ & \ion{Fe}{i} & $3.237$ & $10843.86$ & \ion{Si}{i} & $5.863$ & $-2479$ & $3308$ & $9$ & $146$ & \\
(45) & 52 & $10834.85$ & \ion{Na}{i} & $3.617$ & $10784.56$ & \ion{Si}{i} & $5.964$ & $-3081$ & $4923$ & $9$ & $123$ & \\
(46) & 52 & $10801.36$ & \ion{Cr}{i} & $3.011$ & $10811.12$ & \ion{Mg}{i} & $5.946$ & $-2476$ & $2985$ & $9$ & $136$ & \\
(47) & 52 & $10816.91$ & \ion{Cr}{i} & $3.013$ & $10827.09$ & \ion{Si}{i} & $4.954$ & $-2020$ & $3002$ & $8$ & $131$ & \\ \hline 
\end{tabular}
\end{table*}

\begin{table*}
\caption{Same as \autoref{table:resultY} but for line pairs in \textit{J} band.}
\label{table:resultJ}
\begin{tabular}{ccrcrrcrrrcrr} \hline 
 & & \multicolumn{3}{c}{Low-excitation Line} & \multicolumn{3}{c}{High-excitation Line} & \multicolumn{4}{c}{LDR--$T_{\mathrm{eff}}$ Relation} & \\
ID & Order & \multicolumn{1}{c}{$\lambda $ [\AA ]} & El. & \multicolumn{1}{c}{E.P. [eV]} & \multicolumn{1}{c}{$\lambda $ [\AA ]} & El. & \multicolumn{1}{c}{E.P. [eV]} & \multicolumn{1}{c}{$a$ [K]} & \multicolumn{1}{c}{$b$ [K]} & $N$ & \multicolumn{1}{c}{$\sigma \ \mathrm{[K]}$} & flag \\ \hline 
(48) & 48 & $11638.26$ & \ion{Fe}{i} & $2.176$ & $11640.94$ & \ion{Si}{i} & $6.274$ & $-2932$ & $6052$ & $9$ & $129$ & \\
(49) & 48 & $11667.23$ & \ion{Ti}{i} & $2.345$ & $11681.59$ & \ion{Fe}{i} & $3.547$ & $-983$ & $3668$ & $5$ & $49$ & \\
(50) & 48 & $11797.19$ & \ion{Ti}{i} & $1.430$ & $11793.04$ & \ion{Ca}{i} & $4.535$ & $-1792$ & $4315$ & $8$ & $94$ & \\

(51) & 47 & $11884.08$ & \ion{Fe}{i} & $2.223$ & $11984.20$ & \ion{Si}{i} & $4.930$ & $-4890$ & $4172$ & $9$ & $103$ & \\
(52) & 47 & $11955.95$ & \ion{Ca}{i} & $4.131$ & $12005.40$ & \ion{Fe}{i} & $5.587$ & $-3333$ & $3862$ & $9$ & $142$ & \\
(53) & 47 & $12000.97$ & \ion{Cr}{i} & $3.435$ & $12039.82$ & \ion{Mg}{i} & $5.753$ & $-1954$ & $3329$ & $6$ & $39$ & \\

(54) & 46 & $12105.84$ & \ion{Ca}{i} & $4.554$ & $12178.34$ & \ion{Si}{i} & $6.269$ & $-2366$ & $4660$ & $9$ & $116$ & \\
(55) & 46 & $12255.70$ & \ion{Ti}{i} & $3.921$ & $12133.99$ & \ion{Si}{i} & $5.984$ & $-1971$ & $4202$ & $5$ & $90$ & \\
(56) & 46 & $12190.10$ & \ion{Fe}{i} & $3.635$ & $12175.73$ & \ion{Si}{i} & $6.619$ & $-3204$ & $5735$ & $9$ & $144$ & \\
(57) & 46 & $12267.89$ & \ion{Fe}{i} & $3.274$ & $12283.30$ & \ion{Fe}{i} & $6.169$ & $-1948$ & $4368$ & $6$ & $140$ & \\

(58) & 45 & $12340.48$ & \ion{Fe}{i} & $2.279$ & $12390.15$ & \ion{Si}{i} & $5.082$ & $-2374$ & $4091$ & $9$ & $69$ & 1 \\
(59) & 45 & $12388.37$ & \ion{Ti}{i} & $2.160$ & $12393.07$ & \ion{Fe}{i} & $4.956$ & $-2413$ & $4345$ & $8$ & $114$ & 1 \\
(60) & 45 & $12557.00$ & \ion{Fe}{i} & $2.279$ & $12423.03$ & \ion{Mg}{i} & $5.932$ & $-4690$ & $4993$ & $9$ & $144$ & \\
(61) & 45 & $12532.84$ & \ion{Cr}{i} & $2.709$ & $12457.13$ & \ion{Mg}{i} & $6.431$ & $-1892$ & $5119$ & $9$ & $84$ & \\
(62) & 45 & $12510.52$ & \ion{Fe}{i} & $4.956$ & $12583.92$ & \ion{Si}{i} & $6.616$ & $-1965$ & $4468$ & $9$ & $129$ & \\

(63) & 44 & $12600.28$ & \ion{Ti}{i} & $1.443$ & $12720.15$ & \ion{Co}{i} & $3.530$ & $-1302$ & $4869$ & $8$ & $76$ & \\
(64) & 44 & $12671.10$ & \ion{Ti}{i} & $1.430$ & $12789.45$ & \ion{Fe}{i} & $5.010$ & $-1637$ & $4885$ & $8$ & $68$ & \\
(65) & 44 & $12744.91$ & \ion{Ti}{i} & $2.488$ & $12679.17$ & \ion{Na}{i} & $3.617$ & $-2061$ & $3314$ & $5$ & $145$ & \\
(66) & 44 & $12831.44$ & \ion{Ti}{i} & $1.430$ & $12807.15$ & \ion{Fe}{i} & $3.640$ & $-2806$ & $4808$ & $9$ & $76$ & \\
(67) & 44 & $12821.67$ & \ion{Ti}{i} & $1.460$ & $12808.24$ & \ion{Fe}{i} & $4.988$ & $-3021$ & $5954$ & $8$ & $93$ & \\
(68) & 44 & $12811.48$ & \ion{Ti}{i} & $2.160$ & $12870.04$ & \ion{Mg}{i} & $6.588$ & $-1913$ & $5418$ & $8$ & $121$ & \\
(69) & 44 & $12847.03$ & \ion{Ti}{i} & $1.443$ & $12840.57$ & \ion{Fe}{i} & $4.956$ & $-4468$ & $5627$ & $9$ & $65$ & \\

(70) & 43 & $12910.09$ & \ion{Cr}{i} & $2.708$ & $12896.12$ & \ion{Fe}{i} & $4.913$ & $-3182$ & $5019$ & $9$ & $88$ & \\
(71) & 43 & $12921.81$ & \ion{Cr}{i} & $2.709$ & $12909.07$ & \ion{Ca}{i} & $4.430$ & $-3019$ & $3422$ & $8$ & $61$ & \\
(72) & 43 & $12919.90$ & \ion{Ti}{i} & $2.154$ & $13014.84$ & \ion{Fe}{i} & $5.446$ & $-1428$ & $4822$ & $8$ & $48$ & \\
(73) & 43 & $12927.48$ & \ion{Ti}{i} & $2.154$ & $13134.94$ & \ion{Ca}{i} & $4.451$ & $-1181$ & $3268$ & $5$ & $134$ & \\
(74) & 43 & $12932.31$ & \ion{Ni}{i} & $2.740$ & $12934.67$ & \ion{Fe}{i} & $5.393$ & $-5366$ & $5703$ & $9$ & $122$ & \\
(75) & 43 & $12937.02$ & \ion{Cr}{i} & $2.710$ & $13098.88$ & \ion{Fe}{i} & $5.010$ & $-2800$ & $4811$ & $9$ & $140$ & \\
(76) & 43 & $12975.91$ & \ion{Mn}{i} & $2.888$ & $13029.52$ & \ion{Si}{i} & $6.083$ & $-2655$ & $6276$ & $9$ & $111$ & \\
(77) & 43 & $12987.57$ & \ion{Ti}{i} & $2.506$ & $13147.92$ & \ion{Fe}{i} & $5.393$ & $-1723$ & $3694$ & $6$ & $101$ & \\
(78) & 43 & $13005.36$ & \ion{Ti}{i} & $2.175$ & $13152.74$ & \ion{Si}{i} & $4.920$ & $-1807$ & $3864$ & $6$ & $84$ & \\
(79) & 43 & $13006.68$ & \ion{Fe}{i} & $2.990$ & $13030.92$ & \ion{Si}{i} & $6.079$ & $-2665$ & $4735$ & $9$ & $75$ & \\
(80) & 43 & $13011.90$ & \ion{Ti}{i} & $1.443$ & $13123.41$ & \ion{Al}{i} & $3.143$ & $-1819$ & $3313$ & $8$ & $126$ & \\
(81) & 43 & $13033.55$ & \ion{Ca}{i} & $4.441$ & $13102.06$ & \ion{Si}{i} & $6.083$ & $-3156$ & $4148$ & $8$ & $149$ & \\ \hline 
\end{tabular}
\end{table*}

\section{Discussion}\label{sec:discussion}
It is useful to know what spectral quality is needed to measure the effective temperature based on our LDR method for planning future observations.
Here, we discuss how the precision of $T_{\mathrm{LDR}}$ depends on the S/N of a spectrum and on whether or not we can make the telluric correction.

\begin{figure}
\includegraphics[width=\columnwidth ]{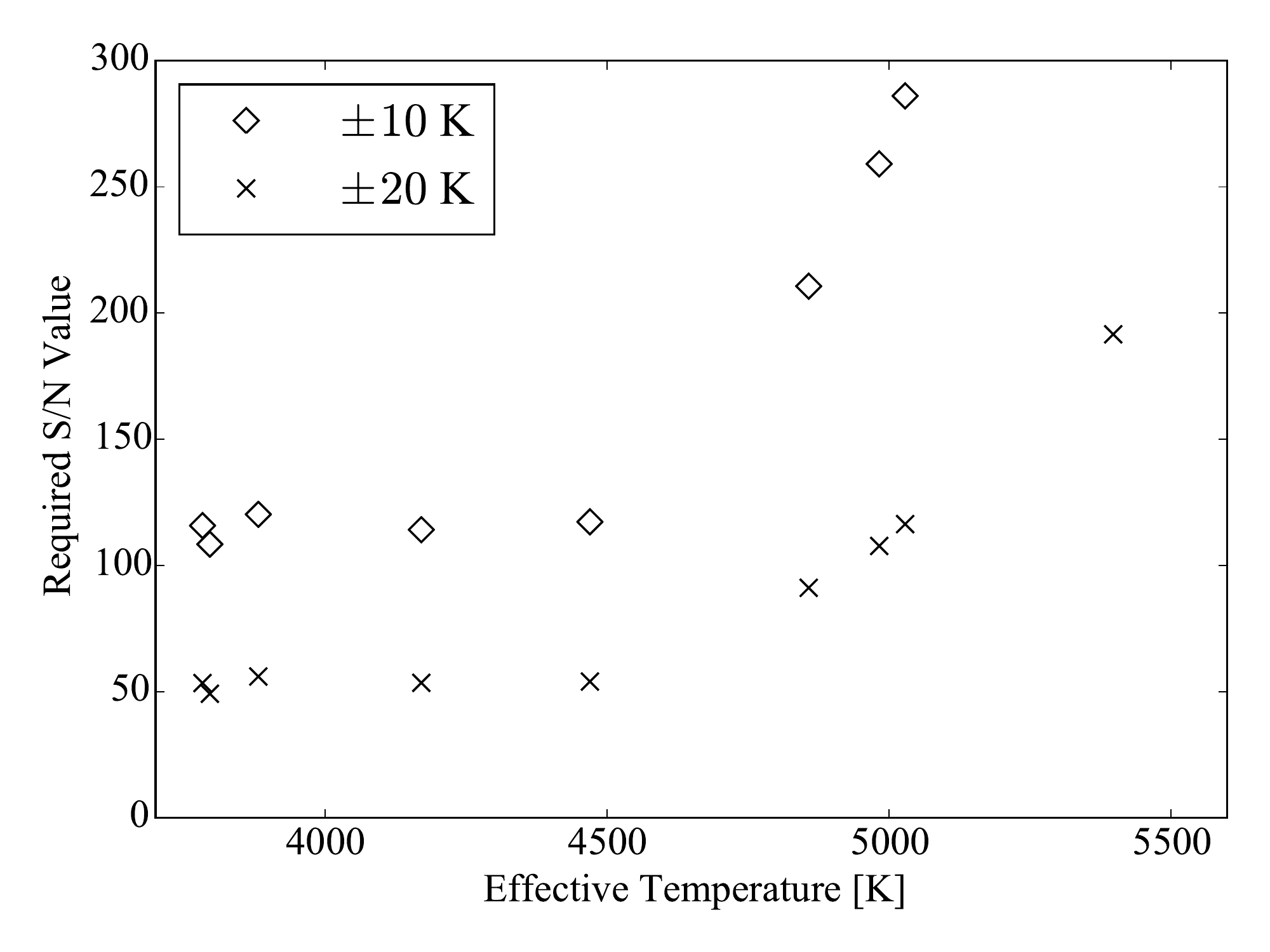}
\caption{The value of S/N required for estimating $T_{\mathrm{LDR}}$ with specific precision: $10\ur{K}$ (open diamond) and $20\ur{K}$ (cross). See more details in the text.}
\label{fig:need_SN}
\end{figure}

First, considering a common S/N fixed for all orders, we calculate a feasible precision by $\Delta {T'_{\mathrm{LDR}}}^{(n)}=\sqrt{1/\sum _{j}(1/\Delta T_{j}^{(n)})^{2}}$, where $\Delta T_{j}^{(n)}$ is the temperature error for each star obtained in the same way described in \autoref{sec:analysis} except that the given S/N is used instead of the S/N obtained from our observed spectrum.
It is straightforward to calculate the $\Delta T'_{\mathrm{LDR}}$ as a function of S/N, and we can estimate with which S/N the precision reaches two target values, $10$ and $20\ur{K}$.
As illustrated in \autoref{fig:need_SN}, in order to estimate the effective temperature of $4000\ur{K}$ within the precision of $10\ur{K}$, for example, one needs to obtain spectra with S/N higher than $100$.
Although required S/N values depend also on the metallicity and the resolution which change the line depths, \autoref{fig:need_SN} can be used as a guideline for estimating the necessary observational times for solar-metal stars approximately using the instrument with the resolution $\sim 28,000$.
The statistical error shall get larger if a limited wavelength coverage allows one to use a smaller number of lines.

Next, we compare the precision which can be achieved with telluric-corrected or uncorrected spectra.
We found that all the metal lines, used in this work, in the 53--55th orders and also a significant number of lines in other orders are free from blending with telluric lines for our objects.
When the LDR method is applied to other stars, however, one needs to take radial velocities into account for considering the blending.
For example, some of the lines in the 53rd to 55th orders could have been blended with telluric lines if our targets were offset by a particular redshift.
For 23 line pairs marked in \autoref{table:resultY} and \autoref{table:resultJ} (21 in \textit{Y} band and two in \textit{J} band), both lines of each pair have no telluric absorption line deeper than $3\%$ with the spectral range corresponding to the redshift within $200\ur{\si{km.s^{-1}}}$.
We considered the absorption depths of a telluric standard spectrum of \textomikron \ Aur taken on 2013 November 30.
\autoref{table:reesttemp_nottre} compares the redetermined effective temperatures obtained with all line pairs in telluric-corrected spectra ($T_{\mathrm{LDR}}$) and the values obtained with the above 23 line pairs in telluric-uncorrected spectra ($T_{\mathrm{LDR}}^{\mathrm{nott}}$).
The error propagation was calculated in the same way as for $T_{\mathrm{LDR}}$ as mentioned above and S/N values in \autoref{table:SN} were used also for telluric-uncorrected spectra.
We found that $T_{\mathrm{LDR}}^{\mathrm{nott}}$ are consistent with $T_{\mathrm{LDR}}$.
Even with significantly fewer line pairs, the precision of $T_{\mathrm{LDR}}^{\mathrm{nott}}$ is still moderately high.
Our LDR method can be used even without the telluric correction with slightly lower precision, which may give efficient access to stellar temperatures with less observational times and analytical works.

\begin{table}
\caption{The literature values of atmospheric parameters, the redetermined effective temperatures using all the available line pairs ($T_{\mathrm{LDR}}$) and the ones based on uncorrected spectra ($T_{\mathrm{LDR}}^{\mathrm{nott}}$). The numbers in round brackets indicate the numbers of line pairs used for the LDR temperatures. The literature $T_{\mathrm{eff}}$ is same as \autoref{table:usedstar}.}
\label{table:reesttemp_nottre}
\begin{tabular}{llll}\hline 
Object & $T_{\mathrm{eff}}\ \mathrm{[K]}$ & $T_{\mathrm{LDR}}\ \mathrm{[K]}$ & $T_{\mathrm{LDR}}^{\mathrm{nott}}\ \mathrm{[K]}$ \\ \hline 
\textepsilon \ Leo & $5398\pm 31$ & $5429\pm 24$ (42) & $5395\pm 46$ (14) \\
\textkappa \ Gem & $5029\pm 47$ & $4982\pm 13$ (60) & $4971\pm 15$ (19) \\
\textepsilon \ Vir & $4983\pm 61$ & $4996\pm 11$ (65) & $5024\pm 19$ (20) \\
Pollux & $4858\pm 60$ & $4829\pm 12$ (73) & $4836\pm 23$ (22) \\
\textmu \ Leo & $4470\pm 40$ & $4434\pm 15$ (80) & $4454\pm 27$ (23) \\
Alphard & $4171\pm 52$ & $4143\pm 9$ (79) & $4169\pm 13$ (23) \\
Aldebaran & $3882\pm 19$ & $3887\pm 6$ (78) & $3860\pm 13$ (21) \\
\textalpha \ Cet & $3796\pm 65$ & $3780\pm 9$ (79) & $3780\pm 14$ (21) \\
\textdelta \ Oph & $3783\pm 20$ & $3812\pm 4$ (79) & $3806\pm 7$ (21) \\
Arcturus & $4286\pm 35$ & $4312\pm 36$ (71) & $4261\pm 72$ (18) \\ \hline 
\end{tabular}
\end{table}

The accuracy of the continuum normalisation is another important factor which may affect the accuracy of the LDR measurements.
For example, \citet{2017A&A...601A..38J} found that offsets between re-normalised continuum levels obtained with different methods are as large as $0.05$ in the worst case for optical lines of Arcturus (see their Fig. 4).
If we add (or subtract) such a large offset to the depth of each line, the LDR temperatures could change by $100\text{--}200\ur{K}$.
However, the effect of the normalisation is expected to be smaller in the LDR method because the lines selected in this work tend to be in relatively uncrowded spectral parts for which the continuum is easily traced.
In addition, the LDR--$T_{\mathrm{eff}}$ relations are based on the empirical calibration, and the line depths are not compared with a theoretical prediction in which the continuum is perfectly defined.
Even if there were some offsets between the true and apparent continuum levels, the LDR values for stars with similar temperatures should be similar.
In fact, the statistics of the LDR--$T_{\mathrm{eff}}$ relations we obtained indicates that the continuum
normalisation has little impact on the precision of our LDRs.
The scatters around individual LDR--$T_{\mathrm{eff}}$ relations are $10\text{--}150\ur{K}$, and they can be explained by statistical errors according to the spectral quality, S/N, and the precision of the reference temperatures.
Nonetheless, when LDRs are measured in spectra with a significantly different resolution, the continuum normalisation may cause systematic temperature offsets of the order of $100\text{--}200\ur{K}$ in the worst cases and the re-calibration with a homogeneous dataset is recommended.

\section{Summary and Conclusion}
We developed the method to estimate the effective temperature using the LDR in \textit{Y} and \textit{J} bands.
This method enables us to estimate the effective temperatures of G, K and M giants with the precision of $\pm 10\ur{K}$ in the best cases using 81 line pairs.
Roughly speaking, the S/N of $100$ is needed to estimate the effective temperature to the precision of $\pm 10\ur{K}$ for a solar-metal star with $T_{\mathrm{eff}}=4000\ur{K}$, while without the telluric correction the error would get two times larger.
Although our set of line pairs is optimized for use with WINERED in terms of the resolution and the wavelength range of the individual orders, this set should also be useful for other instruments.
Re-calibration of the LDR--$T_{\mathrm{eff}}$ relations would be useful for other instruments although LDRs are expected to be insensitive to the spectral resolution under no blending effects of neighbouring lines.

\section*{Acknowledgements}
We acknowledge the useful comments from the referee, Ulrike Heiter.
We are greatful to Masaomi Kinoshita, Kenshi Nakanishi, Tetsuya Nakaoka and Yoshiharu Shinnaka for observing a part of our targets.
We also thank the staff of Koyama Astronomical Observatory for their support during our observation.
This work has been supported by Grants-in-Aid, KAKENHI, from Japan Society for the Promotion of Science (JSPS; Nos. 16684001, 20340042, 21840052 and 26287028) and MEXT Supported Programs for the Strategic Research Foundation at Private Universities (Nos. S0801061 and S1411028).
NK also acknowledges support through the Japan--India Scinece Cooperative Program between 2013 and 2018 under agreement between the JSPS and the Department of Science and Technology (DST) in India.
KF is grateful to KAKENHI (16H07323) Grant-in-Aid for Research Activity Start-up.

\bibliographystyle{mnras}
\bibliography{LDR_taniguchi}

\section*{Supporting Information}
Additional Supporting Information may be found in the online version of this article. \\
\textbf{Figure 3.} The LDR--$T_{\mathrm{eff}}$ plots for all the 81 relations.

\bsp 
\label{lastpage}
\end{document}